\documentclass{pasj00}

\begin{document}
\SetRunningHead{P. T. \.{Z}ycki et al.}{Light bending model}
\Received{2010/03/06}
\Accepted{2010/06/22}

\title{On the light-bending model of X-ray variability of MCG--6-30-15}

%
 \author{%
   Piotr T.\ \textsc{\.{Z}ycki}\altaffilmark{1},
   Ken \textsc{Ebisawa}\altaffilmark{2,3},
   Andrzej \textsc{Nied\'{z}wiecki}\altaffilmark{4}
   and
   Takehiro \textsc{Miyakawa}\altaffilmark{2,3}}
 \altaffiltext{1}{Nicolaus Copernicus Astronomical Center, Bartycka 18, 
   Warsaw, Poland}
 \email{ptz@camk.edu.pl}
 \altaffiltext{2}{The Institute of Space and Astronautical Science, 
   Japan Aerospace Exploration Agency, 3-1-1 Yoshinodai, Chuo, Sagamihara, 
   Kanagawa 229-5210, Japan}
 \altaffiltext{3}{Department of Astronomy, Graduate School of Science,
   University of Tokyo, 7-3-1 Hongo, Bunkyo-ku, Tokyo 113-0033, Japan}
 \altaffiltext{4}{University of {\L}\'{o}d\'{z}, Department of Physics, 
       Pomorska 149/153, 90-236 {\L}\'{o}d\'{z}, Poland}


\maketitle

\begin{abstract}
We apply the light bending model of X-ray variability to {\sl Suzaku\/} data of 
the Seyfert 1 galaxy MCG--6-30-15. We analyze the energy dependence of 
the root mean square 
(rms) variability, and discuss conditions necessary for the model to explain
the characteristic decrease of the source variability around 5-8 keV. 
A model, where the X-ray source moves radially rather than vertically
close to the disk surface, can indeed reproduce the reduced variability
near the energy of the Fe K$_{\alpha}$ line, although the formal fit
quality is poor. The model then predicts the energy spectra, which
can be compared to observational data. The spectra are strongly
reflection dominated, and do {\em not}\/ provide a good fit to {\sl Suzaku\/}
spectral data of the source. The inconsistency of this result with some
previous claims can be traced to our using data in a broader energy band,
where effects of warm absorber in the spectrum cannot be neglected.

\end{abstract}

\section{Introduction}

The Seyfert 1 galaxy MCG--6-30-15 shows the best example of a very broad
Fe K$_{\alpha}$ fluorescent line in its X-ray spectrum (Tanaka et al.\ 1995;
see Reynolds \& Nowak 2003 for review).
According to numerous studies (e.g., Miniutti et al.\ 2007 and 
references therein)
the line is produced well inside the marginally stable orbit for a
Schwarzchild black hole, requiring a (maximally) rotating black hole.
Uncertainties remain, though, as to the robustness of the broadness
of the line, since the source has a very complex ionized absorber
(e.g., Otani et al.\ 1996; Lee et al.\ 2001) and the details of the line 
modelling (e.g.\ its width) are rather sensitive to the description of the 
absorber (Miller, Turner \& Reeves 2008; Miyakawa, Ebisawa \& Inoue 2010).

At the same time, the variability properties of the line are rather
incompatible with a simple geometrical ideas of the central accretion flow. 
The K$_{\alpha}$ line, and the Compton reflection component associated with
it appear to be much less variable than the continuum producing
the reprocessed component. 
This was pointed out by Inoue \& Matsumoto (2001, 2003), and later confirmed
by many authors (e.g., Fabian et al.\ 2002). The line does not follow 
the continuum variability in a simple way, although some changes in the 
line profile/flux are observed (Iwasawa et al.\ 1996).  

In simple geometrical scenarios the line and reflected
continuum should respond to rapid variations of the primary 
continuum driving the reprocessed components. An explicit decomposition 
of the MCG--6-30-15 spectra onto components which can be
interpreted as variable and constant one was done by Miller et al.\ (2008) 
using the Principal Component Analysis method. The constant
component appears to have a shape of a mildly ionized reflection,
while the variable one corresponds to a broadband continuum with
ionized absorption. 

A way of reconciling these two properties of the K$_{\alpha}$ line was 
the so called ``light bending model'' (LBM) by Miniutti \& Fabian (2004).
The idea is that, for a source close to the central black hole, light bending
and other relativistic effects will affect much more the observed primary
emission than the reflected component. Thus, as the position of the
source changes, the observed flux of the primary source will
vary more than the flux of the reflected component.
The model gained popularity despite it being rather ad hoc and having no
obvious relation to the whole body of knowledge on X-ray variability
of accreting black hole systems (see M$^{\rm c}$Hardy 2010 for a recent review
and references).

The properties of the LBM were reanalyzed by Nied\'{z}wiecki \& \.{Z}ycki 
(2008, hereafter NZ08), where it was concluded that in the original 
formulation the model actually failed to explain the data (see also
Nied\'{z}wiecki \& Miyakawa 2010, hereafter NM10).
NZ08 suggested an alternative geometrical scenario, where the source
moves radially (rather than vertically, as in the original formulation),
very close to the disc surface. Then, light bending to the disc plane,
an effect specific to the (extreme) Kerr metric causes the blue peak of 
the line to be approximately constant. The best variant of the model
to explain the lack of variability of the line was that assuming the 
intrinsic source luminosity followed the Keplerian accretion disc emissivity
(see fig.~7 in NZ08). Here, for radii $\le 3\,R_{\rm g}$, where
$R_{\rm g}=G M/c^2$, the line flux is 
almost constant, while the primary source flux changes by a factor of 
$\approx 7$.

In the following paper NM10 extended the model
calculations to include the reflected continuum. The full model was then 
applied the Suzaku rms$(E)$ spectra of MCG--6-30-15.
The rms$(E)$ spectrum is defined as follows,
\[
 {\rm rms}(E) = {1 \over {\langle F(E) \rangle}} 
   \sqrt{\sum_{i}^{N} { {[F_i(E)- \langle F(E) \rangle]^2} \over {N-1}}},
\]
where the sum is over $N$ contributions to flux $F(E)$ at energy $E$, while
${\langle F(E) \rangle}$ is the mean value of the flux. The rms$(E)$ is 
thus a measure of the source variability, as a function of energy, $E$.
The rms$(E)$ spectra could indeed be modelled assuming that the source
is located very close to the black hole, $R=1.6$--$3\,R_{\rm g}$. 
The rms$(E)$ spectra computed assuming that spectra from 3--5 random radii 
within that range contribute to the emission could indeed match the shape 
of the observed rms$(E)$. However, to match the amplitude of the rms$(E)$
the intrinsic source normalization had to be assumed to vary. 
The intrinsic variability deprives the model its major attracting feature, 
namely the explanation of the constancy of the reflected component.
On the other hand it may go some way towards explaining other key features
of observed X-ray variability of Active Galactic Nuclei: 
the power spectrum, hard X-ray time lags, etc.

In this paper we investigate the model properties in some more detail.
Our main purpose is to compute the energy spectra from the model and
compare them with the {\sl Suzaku\/} spectral data of MCG--6-30-15.

\section{The light bending model}

Our computations of the model are described in detail in 
NZ08 and NM10, following the original formulation by Miniutti \& Fabian (2004).
Here we give only a brief summary of its assumptions and properties.
We assume a maximally rotating Kerr black hole and we consider the geometry, 
where the X-ray source moves radially,
very close to the disk surface. The motion is along
the line of constant polar angle, $\theta$, with $H=0.07 R$, 
where $H$ is the
height of the source. The intrinsic luminosity of the source follows
the Keplerian accretion disk emissivity, $L^{\rm PT}(R)$, (Page \& Thorne 1974;
model $S^{\rm PT}$ in NZ08), where
\[
 L^{\rm PT}(R) = {{3 G M {\dot M}} \over 8 \pi R^3} R_{\rm R}(r,a),
\]
where $r=R/R_{\rm g}$, and $R_R(r,a)$ is the relativistic correction
factor, containing the zero-inner torque condition and 
dependent on radial distance $r$ and black hole angular momentum $a$
[see Krolik 1999 for explicit form of $R_{\rm R}(r,a)$]

\begin{figure}
  \begin{center}
    \FigureFile(80mm,80mm){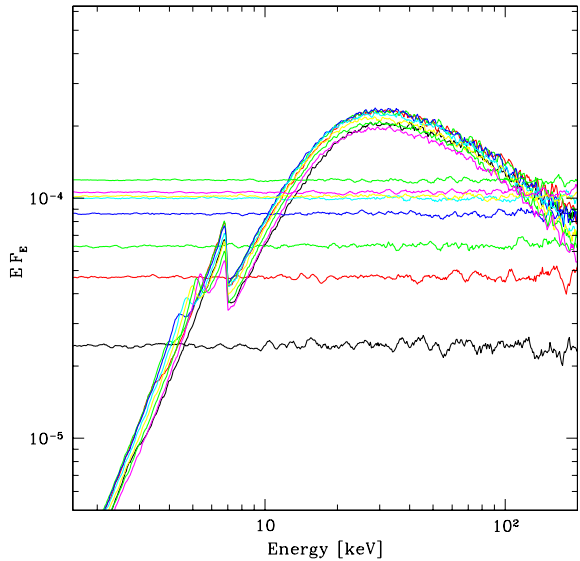}
  \end{center}
  \caption{
The light bending model energy spectra from the $S^{\rm PT}$ model of NZ08, 
i.e.\ assuming that the intrinsic source 
luminosity follows the accretion disk dissipation rate (for a=0.998)
and the source can move along the line of constant polar coordinate with
its height $H=0.07 R$.
Observation (inclination) angle is $i=35^{\circ}$ and the source radial 
positions (corresponding to different curves) are 
$1.6, 1.8, 2, 2.2, 2.4, 2.6, 2.8, 3\,R_{\rm g}$. 
The observed level of the direct source emission (represented by the 
approximately horizontal lines) depends then strongly on the source
position, while the reflected component (and, in consequence the
Fe line) is almost constant (see also fig.~7c in NZ08).
}\label{fig:lbmspectra}
\end{figure}

Then NZ08 show (see fig.~7 there) that, for $R \le 3\, R_{\rm g}$ 
the observed flux of the Fe line 
is  practically constant while the observed flux of the primary continuum
varies by a factor of $\approx 7$. We illustrate this result in 
Fig.~\ref{fig:lbmspectra}, which shows the model spectra for different radii
$\le 3\,R_{\rm g}$. The flux of the reflected component (uniquely related
to the flux of the Fe K$_{\alpha}$ line for a given primary spectral shape and 
reflector ionization) is practically constant, while the primary emission 
changes significantly. 
In the following computations we use the model in the version extended by 
NM10, where
the radial location of the X-ray source is generated from a distribution
$P(r)\propto r^{\delta} $, with $\delta=-1.5$ (sources concentrated towards
the black hole), and the intrinsic luminosity is randomized by being generated
from a gaussian distribution centered at $L^{\rm PT}(R)$, with
standard deviation $\sigma(R)= 0.2L^{\rm PT}(R)$ (note that the randomization
of luminosity is not included in the spectra plotted in 
Fig.~\ref{fig:lbmspectra}).

We note here that the model predicts spectra which are strongly reflection
dominated. This is a necessary consequence of the basic assumptions of 
the light bending model. The major problem then the model will face is
whether such reflection dominated spectra can be compatible with MCG--6-30-15
data. We note that Fabian et al.\ (2002) did suggest that the X-ray spectra
of Seyfert 1 galaxies may be reflection dominated but no fits to data 
of MCG--6-30-15 were presented. Since then, such reflection dominated
spectra have been claimed to be good fits to data from a number of objects,
including MCG--6-30-15.
Most recent analysis of this source was published by
Miniutti et al.\ (2007), where {\sl Suzaku\/} data were analyzed. 
The large value of the amplitude of reflection inferred from the fits,
$\Omega/(2\pi)\approx 4$, is claimed to be consistent with LBM.

\section{Modelling the energy dependence of the r.m.s}
\label{sec:rms}

\begin{figure}
 \begin{center}
  \FigureFile(80mm,80mm){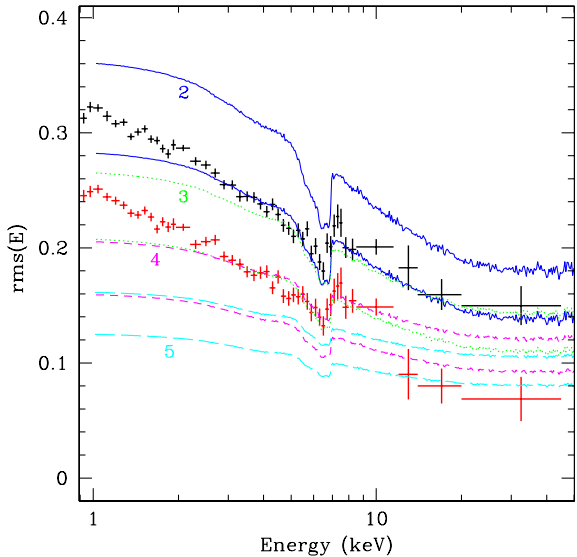}
 \end{center}
 \caption{
The rms$(E)$ dependence from Suzaku data of MCG--6-30-15, together with
model curves. Data points show rms$(E)$ for time bin of 16384 sec (upper set)
and 131072 sec (lower set). Each pair of curves shows a $\pm 1\sigma$ range of
rms$(E)$ from an ensemble of realizations of the model. Numeric labels
show $n_{\rm R}$ -- the number of radii contributing to each spectrum
(see text, Sec~\ref{sec:rms}, for details).}
\label{fig:rmse}
\end{figure}

Results of NZ08 imply that a sequence of spectra from $R \le 3\,R_{\rm g}$ will
produce a rms$(E)$ with a dip around the line energy, since the
amplitude of reflected continuum is the same for all radii 
(Fig.~\ref{fig:lbmspectra}).

Physically, we can expect that a spectrum from a given time interval 
of observation, $T$, will consist  of contributions from  a number of radii, 
the number depending on $T$. 
We treat the number of radii, 
$n_{\rm R}$, as a parameter of the model, since the exact physical processes
responsible for generating active regions are not known. The rms$(E)$ spectrum
is thus computed generating $n_{\rm R}$ values of radius
from the $1.6\,R_{\rm g} \le R \le 3R_{\rm g}$ range, 
adding energy spectra 
from these radii to form one basic spectrum, and then repeating the process 
$N$ times, which corresponds simply to dividing the whole observation into
$N$ intervals. The rms$(E)$ is computed from these $N$ spectra.

A single rms$(E)$ spectrum is just one realization of a random process.
Repeating the above procedure would produce different rms$(E)$.
To estimate the range of possible rms$(E)$ behavior we repeat the 
procedure 1000 times, and plot the resulting  $\pm 1 \sigma$ range of 
rms$(E)$ in Fig.~\ref{fig:rmse}. Since we do not perform formal fitting 
($\chi^2$ minimization) of the model to the rms$(E)$ data, the plot helps 
to estimate the optimum
value of a parameter, in this case $n_{\rm R}$. Specifically, rms$(E)$ computed
with $n_{\rm R}=2$-3 gives the best match to the shorter time scale (16 ksec)
observed rms$(E)$, while $n_{\rm R}=4$-5 might be appropriate for the longer
time scale rms$(E)$, although in the latter case the slope of the model
rms$(E)$ is rather flatter than that of the data.
The normalization of the model rms$(E)$ spectrum is not adjusted to 
best match the data. This is different than the usual practice with
energy spectral fitting, where the model amplitude is adjusted to give lowest
possible $\chi^2$. Therefore the formal $\chi^2$ values computed
from these rms$(E)$ models are rather high, the lowest values of
$\chi^2$ per data point being $\approx 2$. 

\section{Modelling the energy spectra}
\label{sec:spectra}

\begin{figure}
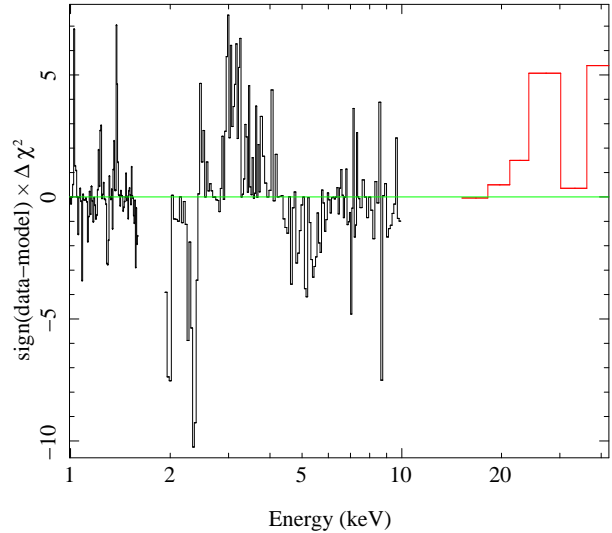

 \begin{center}
 \FigureFile(80mm,80mm){fig3.ps}
 \end{center}
 \caption{
Contributions to $\chi^2$ from fitting the spectral model predicted by 
the LBM matching the rms$(E)$ relation, to the time averaged Suzaku data of 
MCG--6-30-15. The fit is not satisfactory, with $\chi^2_{\nu} \approx 1.5$.
The model spectrum is strongly reflection dominated and extremely 
relativistically smeared, since the source radial position $R\le 3\,R_{\rm g}$.
\label{fig:energychisq}}
\end{figure}

\subsection{Energy spectra predicted by the light bending model}

We now present the energy spectra implied  by the fits of the LBM to 
the rms$(E)$ data and attempt to compare these with the spectral data of
MCG--6-30-15.

Each realization of the rms$(E)$ yields certain mixture of energy spectra 
from a number of radii,
with relative weights resulting from the normalizations generated 
in the process (depending on the radius and the random amplitude around
the $L^{\rm PT}(r)$ value). The total spectrum is then a sum of all these
spectra. The spectra are strongly reflection dominated 
(Fig~\ref{fig:lbmspectra}) with the amplitude of reflection, $\Omega/(2\pi)$ 
reaching a factor of $\approx 5$. 

We fit the January 2006 {\sl Suzaku\/} XIS and PIN time averaged data in the 
range 1-40 keV. Importantly, in this energy range the effects of warm absorber
cannot be neglected, so we fit a model which includes 
description of the absorber. Specifically, we apply a warm absorber model 
constructed from {\sc XSTAR} calculations (Kallman et al.\ 2004). We find
that a double model (low and high ionization) is necessary and sufficient 
to model the data (Sec~\ref{sec:timeaver}; see also Miyakawa et al.\ 2009 
for details of data reduction and warm absorber modelling).

We do preliminary fitting of the spectra from the NZ08 LBM model to the data.
The source's radial position is the most important fitting parameter.
Computations of the reflected component were done only for cold plasma 
($\xi=0$) and for Solar abundances of metals 
(constructing a grid of models is not practical 
due to computing resources required), therefore the model is not as general as
other models available in {\sc XSPEC}. 
The fits are generally poor, with
reduced $\chi^2_{\nu}=1.8-1.9$. Strongest $\chi^2$ residuals appear below 
5 keV.

We construct a more general reflection model by using the results of
computations by Ross \& Fabian (2005). Their model (table model 
{\tt reflion} in {\sc XSPEC}) includes all relevant atomic physics, including 
the Fe line emission (although it is not valid for low ionization, $\xi<30$). 
It was also used in many previous papers (e.g., Miniutti et al.\ 2007).
We now apply relativistic effects by convolving the {\tt reflion} spectrum with 
the transfer function from our LBM calculations for a given radius. Then,
we add a power law continuum normalized so that the amplitude
of the reflected component corresponds to that from the LBM computations.
To achieve this, we use the ratio of the numbers of photons in the primary
and reflected continua above 12 keV (the limit is important to exclude
the effects of ionization). Finally, we sum the spectra over the set
of radii.
The resulting model is thus parameterized by primary power law slope, 
$\Gamma$, ionization parameter, $\xi$, Fe abundance and the inclination angle. 
The amplitude of reflection and relativistic smearing effects 
(i.e.\ the radial emissivity) result form the properties of the LBM and 
are thus fixed for a given set of radii.

This spectral model gives somewhat better fit, but it is still not
satisfactory, with $\chi^2_{\nu} \approx 1.5$ (Fig~\ref{fig:energychisq}). 
Strong
enhancement of reflection drives the primary continuum to be steep,
$\Gamma \approx 2.4$, which seems incompatible with the slope required
for the lower energy range (Fig~\ref{fig:cont}) and, in consequence, 
significant deviations in $\chi^2$ are present at $E<6$ keV. The best
fits are obtained for inclination of $40^{\circ}$ and Fe abundance equal
to Solar. The quality of spectral fits is basically independent of the
model realization, contrary to the quality of the rms$(E)$ fits.


\subsection{Time averaged spectrum of MCG--6-30-15}
\label{sec:timeaver}

According to some earlier studies, the reflection dominated spectra
provide good description of the MCG--6-30-15 data, which fact has been used as 
a supporting argument for the applicability of the LBM.
In particular, Miniutti et al.\ (2007; see also references therein) fitted
the January 2006 Suzaku data with a model comprising cutoff power law
continuum and ionized reflection (XSPEC {\tt reflion} model of Ross \& Fabian
(2005) 
with relativistic smearing modelled with extreme Kerr metric transfer 
function of Laor (1991; XSPEC {\tt kdblur2} model).
Using this model Miniutti et al.\ obtain their best fit with strongly
enhanced reflection, $\Omega/(2\pi)\approx 4$. The reflected component is 
strongly smeared, with inner disk radius of $1.7\,R_{\rm g}$ and illumination
emissivity following very steep, $r^{-4.4}$, dependence up to the  break 
radius of $6\,R_{\rm g}$, beyond which the emissivity is flatter, $r^{-2.5}$.

\begin{figure}
 \begin{center}
 \FigureFile(80mm,80mm){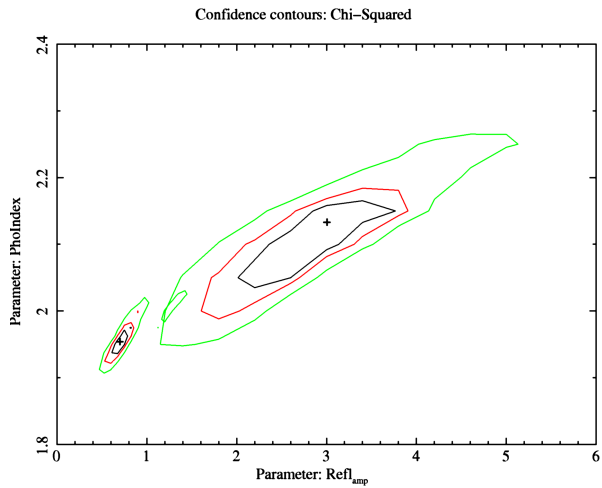}
 \end{center}
 \caption{
The $\chi^2$ contours in the $\Omega/(2\pi)$-$\Gamma$ plane for model fits 
to {\sl Suzaku\/} data of MCG--6-30-15 in two cases: 
data above 3 keV, no warm absorber model (big contours)
and data above 1 keV with a warm absorber model included (small contours).
There is a clear and significant dependence of the amplitude of reflection
on the data range/model components used.
\label{fig:cont}}
\end{figure}

In order to understand then why our results disagree with those earlier ones, 
we compare the results of fitting the same data with the same reflection
model (i.e., the relativistically smeared {\tt reflion} model) in two cases: 
(1) using only data above 
3 keV, assuming that warm absorber does not affect the spectrum (as in Miniutti
et al.\ 2007), and (2)
using the data above 1 keV and attempting to describe the warm absorber.
The warm absorber model, computed with {\sc XSTAR} is prepared as a table
model on a grid of 20$\times$20 values of column density, $N_{\rm H}$, and
ionization parameter, $\xi$ (Miyakawa et al.\ 2009). 
We use two tables (lower and higher ionization),
having checked that a single-$\xi$ model does not provide a good fit, while
adding a third table does not improve the fit. We assume that the 
radial emissivity for line profile computation is the same as in Miniutti et 
al.\ (2007) and we keep it fixed, but we let the inner disk radius and 
inclination to be free. We keep the outer disk radius fixed at 
$400\,R_{\rm g}$.
The Fe abundance is also free to fit, as is obviously the
overall reflection amplitude.

We reproduce the results of Miniutti et al.\ (2007) in the first case, where
only data above 3 keV are used. However, with broader energy band data the
best-fit results are very different. The spectral index is smaller 
(harder spectrum) and the reflection amplitude is much smaller, in fact
less than 1. In Fig.~\ref{fig:cont} we show
the $\chi^2$ contours in the spectral index -- reflection amplitude plane, 
for both considered cases. The inner disk radius, characterizing the
broadening of the line is comparable in both cases. 

Our results mean that, to state it conservatively, the derived parameters 
of the reflection component are strongly model dependent. Fits in broader
energy band yield harder spectrum and, in consequence, much smaller 
amplitude of reflection than fits performed in the $E>3\,$keV energy
band. This is a problem for LBM since it necessarily predicts 
reflection dominated spectra.

It is not our goal here to construct more complex models and to study
the dependence of the reflection parameters of the way how absorption
is modelled. It seems that the solutions of Miller et al.\ (2008) and
Miniutti et al.\ (2007) cover the whole broad range of possibilities.
We simply want to emphasize that, in the context of our
considerations, the support for the LBM depends on modelling
the complex spectrum of MCG--6-30-15.

\section{Discussion and conclusions}
\label{sec:discuss}

The light-bending model is an attempt to reconcile the extreme width
of the Fe line with its lack of variability, first reported in MCG--6-30-15.
It invokes extreme relativistic effects to explain weaker variability
of the reprocessed spectral components compared to the primary components.
Following the original formulation of the model by Miniutti \& Fabian
(2004) we reconsidered the model in greater details (NZ08, NM10)
pointing out some inaccuracies in the original version. Reduced variability
of the Fe line compared to primary continuum is indeed possible, but for
a radially rather than vertically moving source.

Quantitatively, the model is able to reproduce the rms$(E)$ relation, in 
particular the drop near the Fe line energy, but only if some intrinsic 
variability of the X-ray source is assumed (NM10).
That is not necessarily a bad thing since it might go some way 
towards reproducing the usual characteristics of X-ray variability
(e.g., power spectra), which the original model does not address.
One possible test of the model is then to construct the energy spectra
and compare these to the data. The model spectra are obviously strongly 
reflection-dominated and relativistically smeared.

We find that the reflection-dominated spectra do not fit
the data, contrary to previous studies (Miniutti et al.\ 2007 and references 
therein). There
is a significant dependence of the best-fit models on the energy range
of the fitted data. The effects of warm absorber in the spectrum are not
limited to below 3 keV, and so using the data below this energy and 
including a warm absorber model gives very different results compared
to fits without the warm absorber and data above 3 keV. In this respect 
our results can be located between the results of Miller et al.\ (2008)
and Miniutti et al.\ (2007), which can be thought of as absorption--dominated
and reflection--dominated solutions, respectively.
We note that the dependence of reflection parameters on details of the warm 
absorber description applies also to the width of the Fe K$_{\alpha}$ line. 
While this is usually claimed to be extremely broad, data modelling including 
complex warm absorber models (Miller et al.\ 2008; Miyakawa et al.\ 2010) 
yields good descriptions of the broad band data of MCG--6-30-15 without 
extremely broad Fe line.

MCG--6-30-15 is not the only source showing certain 
degeneracy between reflection and absorption in their X-ray spectra.
Similar situation occurs in the X-ray binary XTE J1650-500 where
Miniutti, Fabian \& Miller (2004) 
presented fits of models with extremely broad Fe line, 
while Gierli\'{n}ski \& Done (2006) demonstrated that the
Fe line does not need to be so broad, if complex absorption and reflection
models are applied. Another spectral feature where this kind of ambiguity
is observed are the so called soft X-ray excesses observed in many Active
Galactic Nuclei. 
These can be modelled equally well by absorption and reflection effects 
(Sobolewska \& Done 2007). 
The situation may appear somewhat frustrating, appearing despite years
of data collecting and efforts to understand them. More importantly though,
the implications of the two models are very different and  far-reaching, 
since they touch the base of our interest in X-ray astronomy. Studying effects
of extreme gravity near compact objects was and still is one the main driving 
forces for developing technologies for X-ray observations.

Summarizing, we fit the spectra predicted by the LBM of NZ08 to Suzaku data 
of MCG--6-30-15 and find that the model does {\em not}\/ fit the data,
if the fits use data in the energy range where warm absorber effects are 
important. Detailed results depend on models of warm absorber, with more
complex warm absorber models implying less extreme reflection.

\section*{Acknowledgments} 
 
This work was partly supported by Japan Society for the Promotion of Science
fellowship to PTZ. PTZ thanks ISAS/JAXA for their kind hospitality in 
March 2009.
Partial support for this work came also from the Polish Ministry of Science 
and Higher Education through grant no.\ N203 011 32/1518.


\end{document}